\documentclass[twocolumn,showpacs,showkeys,prd,superscriptaddress]{revtex4-1}

\usepackage{float}
\usepackage{subfigure}
\usepackage{siunitx}
\usepackage{ragged2e}

\usepackage{ulem}

\usepackage{amsmath,amssymb,amsfonts,dsfont,mathrsfs,amsthm}
\usepackage{graphicx}
\usepackage{bbold}
\usepackage{slashed}
\usepackage{centernot}
\usepackage{hyperref}
\usepackage{lmodern}
\usepackage{xcolor}
\usepackage{comment}
\usepackage{epstopdf}
\hypersetup{linktocpage,colorlinks=true,urlcolor=blue,linkcolor=blue,citecolor=red}
\usepackage{feynmf}
\usepackage{array}

\newcommand{\espacio}{\,\,\,\,\,}
 
\newcommand{\abs}[1]{\left|#1\right|}

\newcommand{\Lag}{\mathscr{L}}

\newcommand{\vifh}{\hat{{\bf{e}}}}
\newcommand{\vifhn}[2]{\hat{{\bf{e}}}^{\hat{#1}_{#2}}}
\newcommand{\etah}[2]{{\eta}_{\hat{#1}\hat{#2}}}

\newcommand{\vih}{\hat{e}}
\newcommand{\wfh}{\hat{{\boldsymbol{\omega}}}}

\newcommand{\Rfhn}[4]{\hat{{\boldsymbol{\mathcal{R}}}}^{\hat{#1}_{#2}\hat{#3}_{#4}}}
\newcommand{\Rfhnfree}[4]{\hat{\mathring{{\boldsymbol{\mathcal{R}}}}}^{\hat{#1}_{#2}\hat{#3}_{#4}}}

\newcommand{\Th}{\hat{\mathcal{T}}}
\newcommand{\Tfhn}[2]{\hat{{\boldsymbol{\mathcal{T}}}}^{\hat{#1}_{#2}}}
\newcommand{\hodge}{\star}
\newcommand{\K}{\mathcal{K}}
\newcommand{\Ma}{\mathcal{M}}

\newcommand{\Rh}{\hat{{\mathcal{R}}}}

\newcommand{\Rfh}{\hat{{\boldsymbol{\mathcal{R}}}}}

\newcommand{\Kfhn}[4]{\hat{{\boldsymbol{\mathcal{K}}}}^{\hat{#1}_{#2}\hat{#3}_{#4}}}
\newcommand{\Kfhnud}[4]{\hat{{\boldsymbol{\mathcal{K}}}}^{\hat{#1}_{#2}}{}_{\hat{#3}_{#4}}}

\newcommand{\Df}{{\boldsymbol{D}}}
\newcommand{\df}{{\boldsymbol{d}}}
\newcommand{\gf}{{\boldsymbol{\gamma}}}

\newcommand{\T}{\mathcal{T}}
\newcommand{\free}[1]{\mathring{#1}}
\newcommand{\ghu}[1]{\gamma^{\hat{#1}}}
\newcommand{\ghhhu}[3]{\gamma^{\hat{#1}\hat{#2}\hat{#3}}}
\newcommand{\ghhhd}[3]{\gamma_{\hat{#1}\hat{#2}\hat{#3}}}
\newcommand{\1}{\mathbb{1}}

\newcommand{\diag}{\operatorname{diag}}

\newcommand\dv[1][]{\ensuremath{\mathrm{d}V_{\! #1}}}

\usepackage{tikz}
\usepackage{pgfplots}
\usetikzlibrary{calc,spy}
\usepgfplotslibrary{units}
\pgfplotsset{width=.5\textwidth,compat=newest}
\usetikzlibrary{arrows,shapes,positioning}
\usetikzlibrary{decorations.markings}
\usetikzlibrary{decorations.pathreplacing}
\usetikzlibrary{decorations.pathmorphing}
\tikzstyle directed=[postaction={decorate,decoration={markings,mark=at position .5 with {\arrow{stealth}}}}]
\tikzset{%
  cross/.style={path picture={ 
      \draw[black]
      (path picture bounding box.south east) -- (path picture bounding box.north west) 
      (path picture bounding box.south west) -- (path picture bounding box.north east);
}}}
\tikzset{boson/.style={decorate,decoration={snake}}}

\begin{document}
\title{Torsion in extra dimensions and one-loop observables}

\author{Oscar \surname{Castillo-Felisola}}
\email{o.castillo.felisola@gmail.com} 
\affiliation{Departamento de F\'\i sica, Universidad T\'ecnica Federico Santa Mar\'\i a, Casilla 110-V, Valpara\'\i so, Chile.}
\affiliation{Centro Cient\'ifico Tecnol\'ogico de Valpara\'iso, Valpara\'iso, Chile.}

\author{Crist\'obal \surname{Corral}}
\email{cristobal.corral@postgrado.usm.cl}
\affiliation{Departamento de F\'\i sica, Universidad T\'ecnica Federico Santa Mar\'\i a, Casilla 110-V, Valpara\'\i so, Chile.}

\author{Sergey \surname{Kovalenko}}
\email{sergey.kovalenko@usm.cl}  
\affiliation{Departamento de F\'\i sica, Universidad T\'ecnica Federico Santa Mar\'\i a, Casilla 110-V, Valpara\'\i so, Chile.}
\affiliation{Centro Cient\'ifico Tecnol\'ogico de Valpara\'iso, Valpara\'iso, Chile.}

\author{Iv\'an \surname{Schmidt}}
\email{ivan.schmidt@usm.cl}  
\affiliation{Departamento de F\'\i sica, Universidad T\'ecnica Federico Santa Mar\'\i a, Casilla 110-V, Valpara\'\i so, Chile.}
\affiliation{Centro Cient\'ifico Tecnol\'ogico de Valpara\'iso, Valpara\'iso, Chile.}

\begin{abstract}
  We study gravity with torsion in extra dimensions and derive an effective four-dimensional theory containing four-fermion contact operators at the fundamental scale of quantum gravity in the \si{\TeV} range. These operators may have an impact on the low-energy observables and can manifest themselves or can be constrained  in precision measurements.  We calculate possible contributions of these operators to some observables at the one-loop level. We show that the existing precision data on the lepton decay mode of $Z$boson set limits on the fundamental scale of the gravity and compactification radius, which are more stringent than the limits previously derived in the literature.
\end{abstract}

\maketitle

\section{Introduction}

The Standard Model (SM) of particle physics has proved to be a very successful and predictive framework. Moreover, recently, the ATLAS and CMS experiments discovered a new particle with the approximate mass of \SI{125.6}{\GeV}, which is consistent with the long-awaited last missing element of the SM, the Higgs boson~\cite{Aad:2012tfa,Chatrchyan:2012ufa}.
Nevertheless, this anticipated triumph leaves open many basic questions, which do not allow qualifying the SM as a true fundamental theory, but as a low-energy effective framework. Among these open questions, the huge hierarchy between the electroweak and gravitational scales and the lack of compatibility with gravity are specially important.  Although there exists in the literature many proposals for the solution of the hierarchy problem, it is remarkable that two of the  most popular, supersymmetry and extra dimensions, are intimately related to gravity.

There have been many efforts undertaken in the past toward a deeper understanding of gravity from different perspectives. One of these is to consider gravity as a gauge theory.  In fact, the conventional Einstein-Hilbert Theory (EHT)  of gravity can be interpreted as a gauge theory of the Lorentz group~\cite{PhysRev.101.1597,sciama1964physical,ivanenko1983gauge}.
On the other hand, classification of particles with definite mass and spin is given in flat Minkowski spacetime  in terms of irreducible representations of the Poincar\'e group. Nonetheless, attempts to  construct a gauge theory for the Poincar\'e group in four dimensions have failed~\cite{Kibble:1961ba,PhysRevLett.33.445,PhysRevD.13.3192,PhysRevLett.38.739}.

An alternative view of the EHT is to consider it according to a first-order formalism, where the affine connection and the metric are independent variables.  In  cases where the connection  has a nonvanishing antisymmetric part (called \textit{torsion}), this theory is known as the Einstein-Cartan Theory (ECT) of gravity~\cite{Cartan1922,Cartan1924,Cartan1925}. In the pure gravity case, the torsion's presence does not affect the well-known predictions of the EHT. However, coupling fermionic matter to ECT  gives rise to new interactions of the four-fermion type  (see Refs.~\cite{Kibble:1961ba,Hehl:1976kj,shapiro2002physical,hammond2002torsion}), absent in the EHT, due to the spin-torsion interaction.  

Cosmological or experimental manifestations of these interactions would allow discriminating between these two theories. However, within the minimal torsional generalization in four-dimensional spacetime, the gravity effects in particle interactions at low energies are highly suppressed by the inverse squared of the  Planck mass ($M_{\text{Pl}}\sim\SI{e19}{\GeV}$), making them experimentally unobservable. On the other hand, the key point of the extra-dimensional scenarios~\cite{ArkaniHamed:1998rs,Antoniadis:1998ig,ArkaniHamed:1998nn,Randall:1999ee,Randall:1999vf} for the solution of the hierarchy problem is the reduction of the fundamental gravity scale down to the \si{\TeV} range.  This implies that the gravity-induced  interactions, in particular those which originate from the torsion, become phenomenologically important. In fact, the possibility of observation of the torsion-induced interactions has already been addressed in the literature~\cite{Belyaev:1998ax,Lebedev:2002dp,Kostelecky:2007kx,CastilloFelisola:2012fy,Castillo-Felisola:2014iia}.

In the present paper, we study some phenomenological implications of the torsion-induced four-fermion interactions (TFFIs) in extra dimensions. Specifically, we explore one-loop observables within an effective four-dimensional theory derived from the extra-dimensional one. We focus on the TFFIs contribution to the  $Z$-boson interaction with fermions. Using the existing data on precision tests of the SM~\cite{Altarelli:2004fq,Beringer:1900zz}, we extract a stringent limit on the fundamental scale of gravity. 

The article is organized as follows: In Sec.~\ref{sec:CEF}  we briefly summarize the Cartan-Einstein formalism. In Sec.~\ref{sec:extradim}  we present an extra-dimensional scenario with torsional gravity and derive the corresponding effective four-dimensional theory. In Sec.~\ref{sec:constraints}, data on precision measurement of the $Z$-boson decay rate to the electron-positron pair are used to constraint the parameters of the extra-dimensional theory. 
We conclude with Sec.~\ref{sec:conclusions}, summarizing and discussing our results, as well as some uncovered aspects of TFFIs phenomenology.

\section{\label{sec:CEF} Einstein-Cartan gravity coupled with fermions}

Within the framework of first-order formalism, spin connection and vielbeins are independent fields, and  torsion might not vanish. Including torsion implies the existence of an antisymmetric part of the affine connection
\begin{align}
  \hat{\mathcal{T}}_{\hat{\mu}}{}^{\hat{\lambda}}{}_{\hat{\nu}} \equiv 2\hat{\Gamma}_{[\hat{\mu}}{}^{\hat{\lambda}}{}_{\hat{\nu}]} = \hat{\Gamma}_{\hat{\mu}}{}^{\hat{\lambda}}{}_{\hat{\nu}} - \hat{\Gamma}_{\hat{\nu}}{}^{\hat{\lambda}}{}_{\hat{\mu}},
\end{align}
where hatted  indices denote coordinates on a $D$-dimensional  spacetime ($\Ma$), endowed with a metric $\hat{g}_{\hat{\mu}\hat{\nu}}(x)$,  related to the vielbeins, $\vih^{\hat{a}}_{\hat{\mu}}(x)$, via
\begin{equation}
  \label{metricrelation}
  \hat{g}_{\hat{\mu}\hat{\nu}}(x) = {\eta}_{\hat{a}\hat{b}}\,\vih^{\hat{a}}_{\hat{\mu}}(x)\,\vih^{\hat{b}}_{\hat{\nu}}(x),
\end{equation}
and $\etah{a}{b} = \diag{\left(-,+,\ldots,+\right)}$ is the $D$-dimensional Minkowski metric on the tangent space, $T_x\Ma$.

In  differential forms,  torsion and curvature are defined in terms of the vielbeins and spin connection through the Cartan structure equations,
\begin{align}
  \label{cartantorsion}
  \df\vifhn{a}{} + \wfh^{\hat{a}}{}_{\hat{c}}\wedge\vifhn{c}{} &= \Tfhn{a}{} \equiv \frac{1}{2!}\hat{\T}_{\hat{\mu}}{}^{\hat{a}}{}_{\hat{\nu}}\,dx^{\hat{\mu}}\wedge dx^{\hat{\nu}}, \\
  \label{cartancurvature}
  \df\wfh^{\hat{a}\hat{b}} + \wfh^{\hat{a}}{}_{\hat{c}}\wedge\wfh^{\hat{c}\hat{b}} &= \Rfhn{a}{}{b}{} \equiv \frac{1}{2!}\Rh^{\hat{a}\hat{b}}{}_{\hat{\mu}\hat{\nu}}\,dx^{\hat{\mu}}\wedge dx^{\hat{\nu}},
\end{align}
where $\vifhn{a}{}$ and $\wfh^{\hat{a}}{}_{\hat{c}}$ are the vielbein and spin connection 1-forms, while $\Tfhn{a}{}$ and $\Rfhn{a}{}{b}{}$ are the  torsion and curvature 2-forms, which are used to write the ECT of gravity.

The fermionic matter is introduced via the minimal coupling procedure by  defining the covariant derivative for fermions
\begin{equation}
  \label{covder}
  D_{\hat{a}}\Psi = \hat{E}_{\hat{a}}^{\hat{\mu}}D_{\hat{\mu}}\Psi = \hat{E}_{\hat{a}}^{\hat{\mu}}\left(\partial_{\hat{\mu}}\Psi + \frac{1}{4}(\hat{\omega}_{\hat{\mu}}){}^{\hat{b}\hat{c}}\gamma_{\hat{b}\hat{c}}\Psi\right),
\end{equation}
that keeps invariant the Dirac action under local Lorentz transformation. Above, we used the inverse vielbein, $\hat{E}^\mu_a = \left(\hat{e}^a_\mu\right)^{-1}$, and in general, $\gamma_{a_1 \cdots a_n} = \gamma_{[a_1}\cdots\gamma_{a_n]}$.

In the context of the ECT, we are considering the following action (the bold symbols represent differential forms):
\begin{align}
  \nonumber
  S &= \frac{1}{2\kappa_*^2}\int\frac{\epsilon_{\hat{a}_1\ldots\hat{a}_D}}{(D-2)!}\,\Rfh^{\hat{a}_1\hat{a}_2}\wedge\vifh^{\hat{a}_3}\wedge\ldots\wedge\vifh^{\hat{a}_D} \\
  \nonumber
  &\quad- \sum_{r}\int\bigg(\frac{1}{2}\left(\bar{\Psi}_r\gf\wedge\star\Df\Psi_r - \Df\bar{\Psi}_r\wedge\star\gf\Psi_r\right) \\
  \label{formaction}
  &\qquad + m_r \bar{\Psi}_r\Psi_r\,\frac{\epsilon_{\hat{a}_1\ldots\hat{a}_D}}{D!}\vifh^{\hat{a}_1}\wedge\ldots\wedge\vifh^{\hat{a}_D}\bigg),
\end{align}
where $\kappa_*^2 = 8\pi G_* = M_*^{-(2+n)}$, with $G_*$ and $M_*$  the analog of the Newtonian gravity constant and the reduced Planck mass in $D$ dimensions. In the fermionic sector, $\bar{\Psi}\equiv\imath\Psi^\dagger\gamma^0$ is the Dirac adjoint. The $r$ index indicates the fermion specie (to be clarified below). \mbox{$\gf=\gamma_{\hat{a}}{\bf{e}}^{\hat{a}}$} is the gamma matrix 1-form, and \mbox{$\Df=D_{\hat{a}}{\bf{e}}^{\hat{a}}$}. The symbol $\hodge$ denotes Hodge duality. In principle , some of the fermion masses could be the result of spontaneous symmetry breaking, in particular, the usual electroweak symmetry breaking. As we  comment below, the origin of the fermion masses does not matter for the present analysis.

The equations of motion are found from the principle of least action and yield
\begin{align}
  \label{einsteineom}
  \Rh_{\hat{a}\hat{b}} - \frac{1}{2}\etah{a}{b}\Rh &= \kappa_*^2\,\hat{T}_{\hat{a}\hat{b}} \\
  \label{contorsionfound}
  \Th_{\hat{a}}{}^{\hat{b}}{}_{\hat{c}} = 2\,\hat{\K}_{\hat{a}}{}^{\hat{b}}{}_{\hat{c}} &= -\, \frac{\kappa_*^2}{2}\sum_{r}\bar{\Psi}_r\gamma_{\hat{a}}{}^{\hat{b}}{}_{\hat{c}}\Psi_r,
\end{align}
where $\hat{T}_{\hat{a}\hat{b}}$ is the energy-momentum tensor of fermions and $\hat{\K}_{\hat{a}}{}^{\hat{b}}{}_{\hat{c}}$ is the contorsion tensor. From Eq.~\eqref{cartantorsion}, the spin connection can be split into a torsion-free part plus the contorsion
\begin{align}
  \hat{\omega}_{\hat{\mu}}{}^{\hat{a}\hat{b}} &= \hat{\free{\omega}}_{\hat{\mu}}{}^{\hat{a}\hat{b}} + \hat{\K}_{\hat{\mu}}{}^{\hat{a}\hat{b}}\\
  \intertext{and additionally}
  \Tfhn{a}{} &= {\hat{{\boldsymbol{\mathcal{K}}}}^{\hat{a}}{}_{\hat{b}}}\wedge \hat{{\bf{e}}}^{\hat{b}}  .
\end{align}

The equation of motion for the spin connection [Eq.~\eqref{contorsionfound}] is algebraic, and therefore, it can be substituted into the initial action in order to eliminate the torsion, which acts in this model as an auxiliary field. Toward this end, it is convenient to separate the torsion in  Eq.~\eqref{formaction} by the following decompositions
\begin{align}
  \label{gravdecomp}
  \Rfhn{a}{}{b}{} &= \Rfhnfree{a}{}{b}{} + \free{\Df}\Kfhn{a}{}{b}{} + \Kfhnud{a}{}{c}{}\wedge\Kfhn{c}{}{b}{},\\
  \label{F-decomp}
  \Df\Psi_r \ \, &= \free{\Df}\Psi_r + \frac{1}{4}\Kfhn{a}{}{b}{}\,\gamma_{\hat{a}\hat{b}}\Psi_r,
\end{align}
where as before the circled quantities are torsion free. Integrating out the torsion in Eq.~\eqref{formaction}, one finds the action
\begin{align}\label{4FI}
  S &= \free{S}_\text{grav} + \free{S}_\Psi + \frac{\kappa_* ^2}{32}\sum_{r,s}\int \dv[D]\,(\bar{\Psi}_r\gamma^{\hat{a}\hat{b}\hat{c}}\Psi_r)\,(\bar{\Psi}_s\gamma_{\hat{a}\hat{b}\hat{c}}\Psi_s)
\end{align}
with a contact four-fermion interaction, whose presence is a peculiar prediction of the ECT. The following two features of the TFFIs should be highlighted. First, they conserve lepton flavors due to the flavor blindness of gravity. Second, fermions in these interactions are flavor paired due to the fact that the torsion-fermion interactions arise from the kinetic term of  the Dirac action. 

Another comment is also in order. In the torsional extensions of gravity, the measure is not unique. Here we consider the minimal version of the ECT with the standard measure, although more general theories can be constructed if quadratic terms in the torsion are taken into account (see Ref.~\cite{PhysRevD.18.2730,PhysRevD.22.1915,Baekler2011,Fabbri:2011kq}). Within this framework, gauge fields do not couple to the torsion at the classical level. This is because the one-form gauge fields are singlets of local Lorentz transformations and no covariant derivative is needed (for more details see Ref.~\cite{Benn:1980ea}). Gauge invariance of the theory in this case is also guaranteed~\cite{Hehl:1976kj}.  Demanding gauge invariance for fermions, the covariant derivative from Eq.~\eqref{covder} will be shifted by gauge connections $\hat{V}_{\hat{\mu}}^A$, in order to keep the action invariant under such transformation, and then interactions between fermions and gauge bosons will arise. However, the complete implementation of gauge symmetry, in particular,  the SM gauge symmetry, into the gravity framework is a nontrivial subject of special study, which we do not pretend to in the present paper.

\section{\label{sec:extradim}Extra-dimensions scenario}

A distinctive feature of the ECT is the presence of the four-fermion contact operators discussed in the previous section.  Their manifestation in particle interactions  could provide evidence of the spacetime torsion. However, as seen from Eq.~\eqref{4FI}, these operators are  suppressed  by the inverse of the squared Planck mass, leaving them experimentally unreachable. The situation may dramatically change in extra dimensions, where the fundamental Planck scale can be reduced down to the \si{\TeV} range. Specific extra-dimensional scenarios have been proposed as a solution of the hierarchy problem. The most popular are those with more than two compact extra dimensions, proposed by Arkani-Hamed, Dimopoulos, and Dvali~\cite{ArkaniHamed:1998rs,Antoniadis:1998ig,ArkaniHamed:1998nn} and with only one but large extra dimension of Randall and Sundrum~\cite{Randall:1999ee,Randall:1999vf}. A few generalizations of these scenarios have been considered in Refs.~\cite{DeWolfe:1999cp,Gremm:1999pj,MPS,CastilloFelisola:2004eg}.

We consider the ECT in five-dimensional spacetime with the Randall-Sundrum metric~\cite{Randall:1999ee}
\begin{align}
  \label{RSmetric}
  ds^2 = e^{-2k\abs{y}}\eta_{\mu\nu}dx^\mu\,dx^\nu + dy^2,
\end{align}
where the fifth dimension, $y$, is compactified on an  $S^1/\mathbb{Z}_2$ orbifold, corresponding to the interval \mbox{$0\leq y\leq \pi R$}.

The Kaluza-Klein (KK) decomposition of five-dimensional fermions into chiral four-dimensional ones~\cite{Kehagias:2000au,MPT,CastilloFelisola:2010xh,CastilloFelisola:2012ez}, taking into account for phenomenological reasons only the zero KK modes and using the profiles from Refs.~\cite{Gherghetta:2000qt,Gherghetta:2006ha}, is
\begin{align}
  \Psi_r(x,y) &= f(c_r,y)\bigg(\psi_+(x) + \psi_-(x)\bigg),\\
  \intertext{with}
  f(c_r,y) &= \sqrt{\frac{k\left(1-2c_r\right)}{e^{(1-2c_r)\pi\,kR}-1}}\,e^{(2-c_r)ky},
\end{align}
where  for simplicity we have chosen the same profile for left- and right-handed parts. The $c_i$s coefficients control the localization of fermions. For $c_i>\tfrac{1}{2}$ and $c_i<\tfrac{1}{2}$, they are localized near the Planck and the \si{\TeV} branes, respectively, while for $c=\tfrac{1}{2}$, fermions lie in the bulk.

The Clifford algebra in five dimensions can be constructed using the four-dimensional one. The gamma matrices in five-dimensional spacetime are
\begin{align}
  \ghu{a} = \left(\gamma^a,\gamma^*\right).
\end{align}
With this definition, the product of gamma matrices in Eq.~\eqref{4FI} is 
\begin{align}
  (\ghhhu{a}{b}{c})(\ghhhd{a}{b}{c}) &= 6\left(\gamma_a\gamma^*\right)\left(\gamma^a\gamma^*\right) + 3\left(\gamma^{ab}\gamma^*\right)\left(\gamma_{ab}\gamma^*\right).
\end{align}

Using the chirality condition $\gamma^*\psi_{r\pm} = \pm\psi_{r\pm}$, the TFFIs in Eq.~\eqref{4FI} can be written in the zero-mode approximation as (see Ref.~\cite{Castillo-Felisola:2013jva})
\begin{widetext}
  \begin{align}
    S_{4\text{FI}} &\approx \sum_{r,s}\,\frac{\kappa_{\text{eff}}^2}{32}\,\int d^4x
    \bigg\{6 \left(\bar{\psi}_{r+}\gamma^\mu\psi_{r+} - \bar{\psi}_{r-}\gamma^\mu\psi_{r-} \right) \left(\bar{\psi}_{s+}\gamma^\mu\psi_{s+} - \bar{\psi}_{s-}\gamma^\mu\psi_{s-} \right) \notag\\
    & \qquad +  3\bigg[\left(\bar{\psi}_{r+}\gamma^{\mu\nu}\psi_{r-}\right)\left(\bar{\psi}_{s+}\gamma_{\mu\nu}\psi_{s-}\right) + \left(\bar{\psi}_{r-}\gamma^{\mu\nu}\psi_{r+}\right)\left(\bar{\psi}_{s-}\gamma_{\mu\nu}\psi_{s+}\right)\bigg]\bigg\}, \label{4FI5D}
  \end{align}
  where 
  \begin{align}
    \label{kapparel}
    k_{\text{eff}}^2 \equiv \frac{(2c_r -1)(2c_s - 1)\left(e^{-2\pi\,kR\left(c_r + c_s -1\right)} - 1\right)\,\kappa_*^2\,k}{(4 - 2c_r - 2c_s)\left(e^{\pi\,kR\left(1 - 2c_r\right)}-1\right)\left(e^{\pi\,kR\left(1 - 2c_s\right)}-1\right)}.
  \end{align}
\end{widetext}

Notice that the axial-tensor term in Eq.~\eqref{4FI5D} must be discarded by phenomenological reasons.  This is required by the presence of chiral fermions in the four-dimensional effective theory, leading, as demonstrated in Ref.~\cite{Flachi:2001bj}, to the orbifold boundary condition \mbox{$\pm\gamma^*f_r(y)=f_r(-y)$}.  Since before the dimensional reduction the term $\bar{\Psi}_r\gamma^{\mu\nu}\gamma^*\Psi_r$ is odd under $y\rightarrow-y$,  it must vanish identically~\cite{Lebedev:2002dp}. Thus, in four dimensions, we are left with only the axial-vector torsion-induced interaction in Eq.~\eqref{4FI5D}. We rewrite the corresponding part of the Lagrangian in the form
\begin{align}\label{Lagr-4FI}
  \mathcal{L}_{\text{4FI}} = \sum_{i,j}\,\frac{6 \,\kappa_{\text{eff}}^2}{32}  \mathcal{J}^{\mu}_{(i)}\mathcal{J}_{\mu (j)},
\end{align}
where $i,j$ are the fermion generation induces and the fermion currents can be written in the manifestly 
SM gauge invariant form
\begin{align}
  \mathcal{J}^{\mu}_{(i)} &= \big(\bar{e}_{R}\gamma^{\mu} e_{R} + \bar{u}_{R}\gamma^{\mu} u_{R} +\bar{d}_{R}\gamma^{\mu} d_{R} \notag\\
  & \quad - \bar{L}\gamma^{\mu} L  - \bar{Q}\gamma^{\mu} Q \big)_{(i)}. \label{SMGI-form}
\end{align}
Here $L$ and $Q$ are the left-handed lepton and quark electroweak doublets, while $e_{R}, u_{R}, d_{R}$ are the electroweak singlets. Thus, the torsion generates in four-dimensional theory the SM gauge-invariant four-fermion interactions.  However, as we commented at the end of Sec.~\ref{sec:CEF}, we do not consider the complete implementation of the SM gauge symmetry to the gravity framework in question. Instead, we adopt here the low-energy approach dealing with the broken phase of the SM symmetry, taking into account only the known SM fermions specified in Eq.~\eqref{SMGI-form} with measured values of their masses and participating in the SM interactions plus the torsion-induced interactions in Eq.~\eqref{Lagr-4FI}.

Using the definitions $\kappa_{\text{eff}}^2 \equiv M_{\text{Pl}}^{-2}$ and $\kappa_*^2 \equiv M_*^{-3}$ and the stabilization value $kR\sim10$ (see Ref.~\cite{Goldberger:1999uk}), we obtain 
\begin{align}
  M_{\text{Pl}}^2 \approx 
  \begin{cases}    
    \num{5e-27} \dfrac{M_*^3}{k};  & c_r\simeq c_s\simeq 0 \\[2ex]
    \num{e-24} \dfrac{M_*^3}{k};  & c_r\simeq c_s\simeq 1/2 \\[2ex]
    \num{e-2}\dfrac{M_*^3}{k};  & c_r\simeq c_s\simeq 1.
  \end{cases}
\end{align}

Let us note that in the following the effects of curvature in the effective theory in four dimensions are ignored, due to the fact that the Universe is essentially flat (cf. Ref.~\cite{Larson:2010gs}). This fact has been used before in Refs.~\cite{Carroll:1994dq,Belyaev:1998ax,Kostelecky:2007kx} and allows us to discriminate between the EHT and the ECT of gravity. Moreover,  this  condition is compatible with the independence of the Riemaniann curvature and torsion.

\section{\label{sec:constraints}Constraints from precision measurements of $Z$-boson decay}
Here we analyze the contribution of the torsion-induced interactions to the one-loop form factors of  the gauge bosons trilinear couplings to leptons. 
Considering that the gauge sector of the SM is torsion free, the only effect of torsion comes through the four-fermion contact 
\mbox{(axial-vector)$\otimes$(axial-vector)} terms  in Eq.~\eqref{4FI5D}. 

The neutral gauge boson-fermion vertex
\begin{align}
  \label{feyndiagram}
  \vcenter{\hbox{\includegraphics[height=15ex]{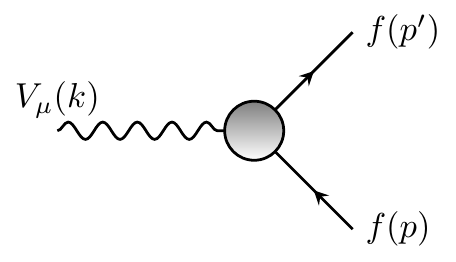}}}
  =\imath e \, V_\mu(k) J^\mu(p,p'),
\end{align}
where $V_\mu = \{\gamma_\mu, Z^0_\mu\}$ can be generically parametrized in terms of the fermion neutral current $J^{\mu}$ form factors $F_{I}$ as
\begin{widetext}
  \begin{align}
    \label{current}
    J^\mu(p,p') &\equiv \bar{u}(p')\Bigg[\gamma^\mu\,F_V(k^2) +F_A(k^2)\gamma^\mu\gamma^* + i\frac{\sigma^{\mu\nu}\,k_\nu}{2\,m_f}F_M(k^2) + F_D(k^2)\frac{1}{2m_f}\sigma^{\mu\nu}\gamma^* k_\nu\Bigg]v(p) .
  \end{align}
\end{widetext}
We decompose the form factors into the tree-level and one-loop contributions $F_i(k^2) = F_i^{\text{tree}} + \delta F_i(k^2)$.  The tree-level part $F_i^{\text{tree}}$ corresponds to the usual SM couplings of photon or $Z$ boson to the fermions, while the one-loop term $\delta F_i(k^2)$ may also receive contributions from beyond the SM interactions. In our case, they are the TFFIs [also known as Eq.~\eqref{4FI5D}]. 

The corresponding one-loop calculations have been carried out  in Ref.~\cite{GonzalezGarcia:1998ay} for the Lagrangian
\begin{align}
  \nonumber
  \Lag_{\text{V}} &= \eta_V\,\frac{g^2}{\Lambda^2}\left[\psi_r\gamma_\mu\left(V_V - A_V\gamma^*\right)\psi_r\right] \\ 
  \label{lagvec}
  &\espacio\espacio\espacio\times \left[\psi_s\gamma^\mu\left(V_V - A_V\gamma^*\right)\psi_s\right],
\end{align}
where $r$ and $s$ denote flavor indices as in the previous sections. We use the results of Ref.~\cite{GonzalezGarcia:1998ay} taking into account both $s$- and $t$-channel contributions with external electrons and with all of the possible particles running in the loop.  Following Ref.~\cite{GonzalezGarcia:1998ay}, we used the normalization $g^2/4\pi = 1$.

Let us note that if $J^\mu(p,p')$ in Eq.~\eqref{current} is coupled to the photon field, the only nonvanishing form factor would be $F_V^\gamma$, due to the absence of (axial) tensor interactions forbidden in  Eq.~\eqref{4FI5D} by the orbifold condition in the RS scenario [see the comment after Eq.~\eqref{kapparel}]. Thus, there are no phenomenologically interesting torsion contributions to the fermionic anomalous magnetic and electric dipole moments.

Then we focus on the $Z^0$ boson coupling to electrons and evaluate the torsion contribution to the corresponding form factors at $Z^0$ pole. The nonvanishing contributions are
\begin{align}
  \label{FVZ}
  \delta F_V^Z(M_Z^2) &= \num{28.7} \, \left(\frac{\si{\GeV}^2}{\Lambda^2}\right)\,\ln\left(\frac{\Lambda^2}{M_Z^2}\right), \\
  \label{FAZ}
  \delta F_A^Z(M_Z^2) &= -\num{3.43e4} \, \left(\frac{\si{\GeV}^2}{\Lambda^2}\right)\,\ln\left(\frac{\Lambda^2}{M_Z^2}\right).
\end{align}
We take the renormalization scale equal to the $Z$-boson mass  $\mu = M_Z$.

With these results, we are ready to calculate the four-fermion torsion contribution to one of the best experimentally studied quantities,  the decay width of $Z$-boson to the electron-positron pair. We decompose the theoretical value of this observable into two parts
\begin{align}
  \label{Gdecomp}
  \Gamma_{\text{th}}\left(Z^0\rightarrow e^+\,e^-\right) &= \Gamma_{\text{SM}} + \delta\Gamma_{\text{4FI}} ,
\end{align}
where 
\begin{eqnarray}
  \label{SM-prediction}
  \Gamma_{\text{SM}} = \SI{84.00 \pm 0.01}{\MeV} 
\end{eqnarray}
is the theoretical prediction of the SM (cf. Ref.~\cite{Beringer:1900zz}) and the four-fermion contribution is given by
\begin{align}
  \nonumber
  \delta\Gamma_{\text{4FI}} &= -\frac{\alpha M_Z}{6\,s_Wc_W}\Bigg[(1 - 4s_W^2)\delta F_V^Z(M_Z^2) + \delta F_A^Z(M_Z^2)\Bigg].
\end{align}
Substituting expressions shown in Eqs.~\eqref{FVZ} and \eqref{FAZ} we find
\begin{align}
  \label{deltagammateo}
  \delta\Gamma_{\text{4FI}} &= \SI{9.87e6}{\MeV}\;\left(\frac{\si{\GeV}}{\Lambda}\right)^2\,\ln\left(\frac{\Lambda^2}{M_Z^2}\right).
\end{align}

The best updated experimental value of $Z^0$ boson decay width  into the electron-positron pair is~\cite{Beringer:1900zz}
\begin{align}
  \label{deltagammaexp}
  \Gamma_{\text{exp}} &= \SI{83.984 \pm 0.086}{\MeV} .
\end{align}
Below we denote the standard experimental deviation from the best fit value as $\Delta_{\text{exp}} = 0.086$ MeV.
We require that the SM contribution taken together with the torsion one be compatible with the experimental data 
\eqref{deltagammaexp}. This leads to the condition
\begin{align}
  \label{compatib}
  \left| \Gamma_{\text{th}} - \Gamma_{\text{exp}}\right|\leq \alpha \Delta_{\text{exp}},
\end{align}
with the statistical factor $\alpha = 1.64$ for the 95\% C.L. limits (cf. Ref.~\cite{Beringer:1900zz}).   Here $\Gamma_{\text{th}}$ is defined in Eq.~\eqref{Gdecomp}.

\begin{center}
  \begin{figure}[H]
    \includegraphics[width=.45\textwidth]{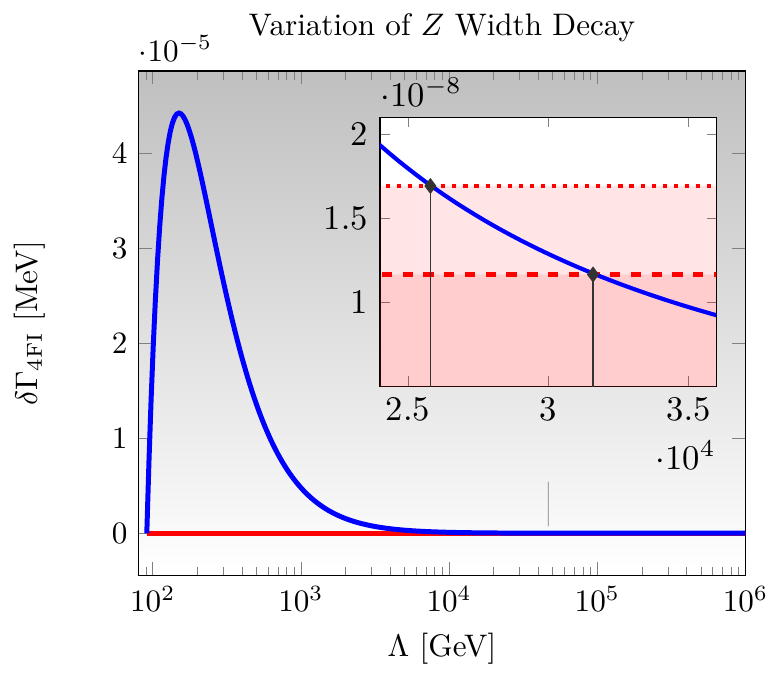}
    \caption{The solid line shows the dependence of  the four-fermion contribution $\delta\Gamma_{\text{4FI}}$ to the $Z^0$ boson decay width on the scale $\Lambda$. The dashed and dotted horizontal lines depict the uncertainty of the SM prediction in Eq.~\eqref{SM-prediction}, and their intersection with the solid curve gives bounds on $\Lambda$, shown as  vertical straight lines on the zoomed part of the plot.}
    \label{Fig:plot}
  \end{figure}
\end{center}

In Fig.~\ref{Fig:plot}, the solid line shows the dependence of  the four-fermion contribution $\delta\Gamma_{\text{4FI}}$ to the $Z^0$-boson decay width on the scale $\Lambda$. 

Solving Eq.~\eqref{compatib}, we find
\begin{align}
  \Lambda &\ge 31.6\,(25.8)\,\SI{}{\TeV} \quad \text{at 95\%C.L.},
\end{align}
depending on which sign is taken for the uncertainty of the SM prediction in Eq.~\eqref{SM-prediction}. These two options are depicted in the figure by the dashed and dotted horizontal lines, respectively.

Comparing with the  coupling of the contact interaction  in Eq.~\eqref{Lagr-4FI} with the one in Eq.~\eqref{lagvec},
\begin{align}
  \frac{6}{32}\,\kappa_{\text{eff}}^2 \longleftrightarrow \frac{1}{\Lambda^2}.
\end{align}
To solve the hierarchy problem in the RS scenario, the value $M_*\sim\SI{1}{\TeV}$ has been fixed. Using the stabilization value $k R \sim 10$ (see Ref.~\cite{Goldberger:1999uk}), one obtains the following 95\% C.L. limits on the compactification radius for different fermion localization values,
\begin{align}
  \label{rconst}
  R \lesssim 
  \begin{cases}
    7.5\,(4.9)\;\SI{e10}{m}; & c_i\simeq 0 \\
    3.7\,(2.5)\;\SI{e8}{m}; & c_i\simeq 1/2 \\
    3.7\,(2.5)\;\SI{e-14}{m}; & c_i\simeq 1 ,
  \end{cases}
\end{align}
where the strongest limit comes from fermions localized near to the Planck brane, which was expected due the enhancement of the gravitational scale close to this brane.

\section{\label{sec:conclusions}Conclusions}

In this paper, we have considered Dirac fermions coupled with the (minimal) ECT of gravity. Within this framework, a four-fermion contact interaction arises, which preserves lepton number and fermions are paired by flavor. For this minimal generalization of the EHT of gravity, the new fermion interaction is suppressed by the gravitational scale, which in four dimensions is the Planck mass, $M_{\text{Pl}}\sim\SI{e19}{\GeV}$.

In order to circumvent this suppression, the gravitational scale should be much lower, not too far from the electroweak scale. This is also suggested by the arguments of radiative stability of the Higgs boson mass. We considered the Randall-Sundrum scenario, which does not suffer of the problems of hierarchical scales. It suggests the existence of one large extra dimension compactified on a $S^1/\mathbb{Z}_2$ orbifold, and a fundamental gravitational scale ($M_*$), which gives rise to an effective (exponentially enhanced) Planck mass through a dimensional reduction.  We implemented the torsional gravity into this scenario. 

The four-fermion interaction in five-dimensions yields both axial-vector and (axial) tensor interactions. Interestingly, the orbifold structure of the extra dimension in RS scenarios requires vanishing the (axial) tensor terms, for the effective four-dimensional theory to contain chiral fermions. For this reason, there are no phenomenologically meaningful torsion contributions to the fermionic anomalous magnetic and electric dipole moments.

On the other hand the remaining \mbox{(axial-vector)$\otimes$(axial-vector)} torsion-induced interactions give a contribution to the width of  $Z^{0}\rightarrow e^{+}e^{-}$, which allowed us to extract from the existing experimental data upper limits on the compactification radius in the RS setup. These limits shown in Eq.~\eqref{rconst} are more stringent then the corresponding limits previously derived in the literature  (see, for instance,  Refs.~\cite{CastilloFelisola:2012fy,Castillo-Felisola:2014iia} and references therein). 

Note that these limits are not going to change even after the complete implementation of the SM gauge symmetry in the ECT gravity framework, not considered in the present work. This is because the only phenomenologically relevant remnant of of the ECT,  the TFFI in Eq.~\eqref{Lagr-4FI}, is already SM gauge invariant. 

\section*{Acknowledgment}

We thank A. Toloza, J. Zanelli, and V. Lyubovitskij for fruitful discussions. This work was supported by Conicyt (Chile) under Grant No. 21130179 and Fondecyt (Chile) under Grants No. 1100582 and No. 1100287.

C. C.   thanks the University of T\"ubingen for hospitality during the completion of this work.

%

\end{document}